\documentclass[twocolumn,superscriptaddress,pre]{revtex4-1}
\usepackage{amsmath, amsfonts, amssymb, graphicx, color}
\graphicspath{{./figures/}}
 
\def\({\left(}
\def\){\right)}                       
\def\[{\left[}
\def\]{\right]}

\newcommand{\<}{\langle}

\newcommand{\av}[1]{\langle{#1}\rangle{}}
\newcommand{\abs}[1]{|#1|}
\newcommand{\beq}{\begin{equation}}
\newcommand{\eeq}{\end{equation}}
\newcommand{\bea}{\begin{align}}
\newcommand{\eea}{\end{align}}

\newcommand{\bx}{ \mathbf{x}}
\newcommand{\by}{ \mathbf{y}}

\renewcommand{\phi}{\varphi}

\newcommand{\ch}{}

\newcommand{\equal}{These authors contributed equally.}

\begin{document}

\author{Lauritz Hahn}
\affiliation{
  Laboratoire de Physique de l'\'Ecole normale sup\'erieure, CNRS, PSL University,\\
  Sorbonne Universit\'e, and Universit\'e Paris Cit\'e, Paris, France}
\author{Aleksandra~M.~Walczak}
\thanks{\equal}
\affiliation{
  Laboratoire de Physique de l'\'Ecole normale sup\'erieure, CNRS, PSL University,\\
  Sorbonne Universit\'e, and Universit\'e Paris Cit\'e, Paris, France}
\author{Thierry~Mora}
\thanks{\equal}
\affiliation{
  Laboratoire de Physique de l'\'Ecole normale sup\'erieure, CNRS, PSL University,\\
  Sorbonne Universit\'e, and Universit\'e Paris Cit\'e, Paris, France}

\title{Dynamical information synergy in biochemical signaling networks}

\begin{abstract}
Biological cells encode information about their environment through biochemical signaling
networks that control their internal state and response. This information is often
encoded in the dynamical patterns of
the signaling molecules, rather than just their instantaneous concentrations.
Here, we analytically calculate the information
contained in these dynamics for a number of paradigmatic cases in the
linear regime, for both static and time-dependent
input signals. When considering oscillatory output dynamics, we report
the emergence of
synergy between successive measurements, meaning that the joint information in
two measurements exceeds the sum of the individual information. We
extend our analysis numerically beyond the scope of linear input encoding
to reveal synergetic effects in the cases of
frequency or damping modulation, both of which are relevant to
classical biochemical signaling systems.
\end{abstract}

\maketitle


To react and adapt to varying internal and external conditions, cells use networks of signaling proteins to convey and process information. 
Recent experiments have shown that these networks often show complex dynamical behavior, such as relaxation to a steady state, pulses, oscillations or bistable switches \cite{Kholodenko2006,Purvis2013,Potter2017}. Given a specific regulatory network topology, different stimuli can produce distinct dynamical responses by messenger molecules. For example, the transcription factor nuclear factor kappa-B (NF-$\kappa$B) exhibits damped oscillations in cells stimulated with tumor necrosis factor-$\alpha$ (TNF$\alpha$) \cite{Hoffmann2002,Covert2005} whereas stimulation with bacterial lipopolysaccharide (LPS) leads to a single, prolonged wave \cite{Covert2005}. These distinct responses help to explain how NF-$\kappa$B can be involved in such diverse processes as inflammatory response, cell differentiation, cell proliferation, apoptosis, and more \cite{Hoffmann2002,Werner2005}. This has led to the hypothesis that cells use the temporal dynamics of signaling molecules to transmit information about both identity and intensity of stimuli \cite{Kholodenko2006,Purvis2013}.

Information theory has been a useful tool for quantifying the information flow in biochemical networks \cite{Margolin2006,Tkacik2008c,Cheong2011,Tkacik2011}. While its application is often restricted to static measurements of the output signal, recent experimental studies have used information theory to quantify the reliability of signal transmission in biochemical networks by estimating the mutual information (MI) between input stimuli and the dynamical responses of signaling molecules such as NF-$\kappa$B \cite{Cheong2011,Selimkhanov2014,Tang2021}, extracellular signal-regulated kinase (ERK) \cite{Selimkhanov2014,Dessauges2022}, calcium ions (Ca$^{2+}$) \cite{Selimkhanov2014,Potter2017}, or nuclear translocation of transcription factors \cite{Granados2018}. These results indicate that information transmission is increased when considering the temporal dynamics as compared to a static scenario.
Concomitantly, progress has been made in calculating analytically the mutual information between input stimuli and dynamical responses for time traces of infinite lengths in simple linearized models of biochemical signaling \cite{Tostevin2009,Tostevin2010}. However, computations of information for trajectories of finite lengths have {\ch mostly} been limited to numerical investigation {\ch\cite{Duso2019a,Reinhardt2022} (sometimes aided by analytical approximations \cite{Moor2023})}, despite their relevance for cells that make quick decisions in response to external cues.
Here we develop a framework for computing analytically the mutual information between the input and the time trace of the output signal, using a linear approximation.
{\ch Focusing on the onset of a constant input, we}  demonstrate the existence of regimes in which information transmission is synergetic, i.e. the information contained in two time points jointly is larger than the sum of the information contained in the individual time points.

\begin{figure}[t!]
\centering
  \includegraphics[width=\linewidth]{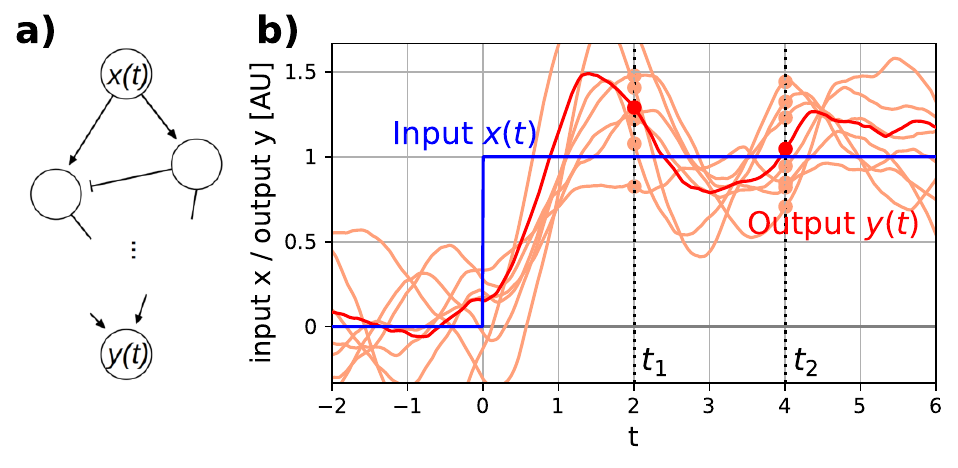}
\caption{\textbf{Calculating mutual information in time traces.} (a) We consider an input signal, either static or dynamic, that is transmitted through a biochemical network and thus produces a time-dependent output signal whose dynamic patterns may encode information about the input. (b) To evaluate the information transmission, we sample the input/output trajectories at a given number of time points and compute the mutual information. In certain cases this is possible analytically, otherwise numerical estimators can be used. 
}
\label{fig1}
\end{figure}

We consider a biochemical network with one input species $X$ and one output species $Y$ (Fig.~\ref{fig1}a). Our goal is to calculate the mutual information between a time trace of the input concentration, denoted by vector $\mathbf{x}\equiv (x_{t'_1},\ldots,x_{t'_m})$, and a set of measurements of the output concentration $\mathbf{y}\equiv (y_{t_1},\ldots,y_{t_n})$: $I(\mathbf{x}; \mathbf{y})=\int d\bx d\by p(\bx,\by)\log p(\bx,\by)/p(\bx)p(\by)$  (Fig.~\ref{fig1}b).

In general, changes in the output species $Y$ are determined by the history of $x$ and $y$ themselves. We start by assuming that this dependency is instantaneous, with no delays, so that increments of $y$ happen with a rate depending on $x_t$ and $y_t$ only (we will relax that assumption later). We also assume $y\gg 1$, so that we can use the small-noise approximation and describe the evolution of $y$ through the stochastic differential equation:
\beq
\dot y_t=f(x_t,y_t)+\sqrt{2D(x_t,y_t)}\eta_t,
\eeq
where $f$ is an arbitrary regulation function that subsumes all the details of the regulation network between $Y$ and $X$, and $\eta_t$ a unitary Gaussian white noise.
Linearizing these dynamics around $(x^*,y^*)$, we can write:
\beq
\dot{y}_t=\tau^{-1}\left(x- y_t \right)+ \sqrt{2D}\eta_t,\label{eq_y_over_EOM}
\eeq
where $x$ and $y$ were shifted and rescaled without losing generality so that $(x^*,y^*)=(0,0)$, and $\partial_yf=-\partial_xf$.

We start by considering the situation in which the input changes from 0 to a value $X\sim \mathcal{N}(0,\sigma_x)$ at $t=0$, corresponding to the sudden activation of the signaling pathway, e.g. due to some environmental change. Eq.~\eqref{eq_y_over_EOM} may be integrated exactly, so that $y_t$ conditioned on $x$ is normally distributed:
\beq
p(y_t|x)=\frac{1}{Z}\exp\left[-\frac{(y_t-x(1-e^{-t/\tau}))^2}{2D\tau(1-e^{-2t/\tau})}\right],
\eeq
where we have assumed $y(0)=0$.
Since $x$ is also normally distributed, $(x,\mathbf{y})$ is distributed as a multivariate Gaussian, and all mutual information values may be calculated exactly (see Appendix~\ref{Gaussian}). The information $I(X;Y_{t_1})$ carried by a single measurement of $y$ at time $t_1$ can be written as a function of the signal-to-noise (SNR) ratio $S(X;Y_{t_1})$ (Appendix~\ref{AppendixStaticX}):
\begin{align}
  I&=\frac{1}{2}\log(1+S),\label{eq:SNR}\\
  S(X;Y_{t_1})&=\frac{\sigma_x^2}{D\tau} \frac{1-e^{-t_1/\tau}}{1+e^{-t_1/\tau}}. \label{eq_ov_MIstatic}
\end{align}
Typically, $I(X;Y_{t_1})$ is reported in the $t_1\rightarrow \infty$ limit, where it is maximal, yet cells can rarely wait that long for a readout. However, a cell is not limited to one measurement of $y$: it can make multiple measurements, or exploit the information contained in the whole output trajectory.

To calculate the mutual information between a static input $X$ and an interval of the output trajectory $\mathbf{y}=\{y_t\}_{t\in [t_1,t_1+T]}$, we use Bayes's law to write the posterior probability of $x$ given the history of $y$ as:
\begin{equation}
  p(x|\mathbf y)=\frac{p(x)p(y_{t_1}|x)}{p(\mathbf y)}\prod_{t={t_1,t_1+\delta t,\ldots}} p(y_{t+\delta t}|y_{t},x) \label{eq_Bayes}.
\end{equation}
The logarithm of each term of this product is quadratic in $x$, meaning that the posterior is Gaussian. Collecting the terms in $x^2$ and taking the $\delta t\to 0$ limit gives us the inverse of the posterior variance $\mathrm{Var}(X|\mathbf{y})$,
from which we deduce the mutual information $I(X;{\bf Y})$ contained in the entire trajectory as \eqref{eq:SNR} with
\beq
S(X;{\bf Y})=\frac{\sigma^2_x }{D\tau}\frac{1-e^{-t_1/\tau}}{1+e^{-t_1/\tau}}+\frac{\sigma^2_x T}{2 D \tau^2}. 
\label{eq_MI_Gauss_over}
\eeq
This SNR is the sum of the SNR given by a single measurement \eqref{eq_ov_MIstatic}, and that provided by an effective number $T/2\tau$ of additional {\it independent} measurements that the trace provides. As usual when combining several measurements, the SNR grows linearly and the MI logarithmically according to a law of diminishing returns, meaning that these measurements are redundant, making synergy between them impossible with these memory-less dynamics.
 
The model of Eq.~\ref{eq_y_over_EOM} cannot describe oscillatory behavior, which is observed in several well-studied systems such as the above-mentioned NF-$\kappa$B, ERK and Ca$^{2+}$, but also other transcription factors such as p53 \cite{Hamstra2006, Batchelor2011}, Crz1 \cite{Cai2008} or Msn2 \cite{Garmendia-Torres2007,Hao2012}. To account for this behaviour, we can consider linearized second-order dynamics, which take the form of an underdamped oscillator under external forcing:
\beq
\ddot{y}_t = -\gamma \dot y_t-\omega_0^2(y_t-x)+{\gamma\sqrt{2D}}\eta_t, \label{eq_un_EOM}
\eeq
where $\gamma$ is the damping coefficient, $\omega_0^2=\gamma/\tau$, and $\Omega =\sqrt{\omega_0^2-\gamma^2/4}$ the natural frequency of the oscillator. Model \eqref{eq_y_over_EOM} corresponds to the overdamped limit $\gamma\to\infty$.

\begin{figure}[t!]
\centering
  \includegraphics[width=1\linewidth]{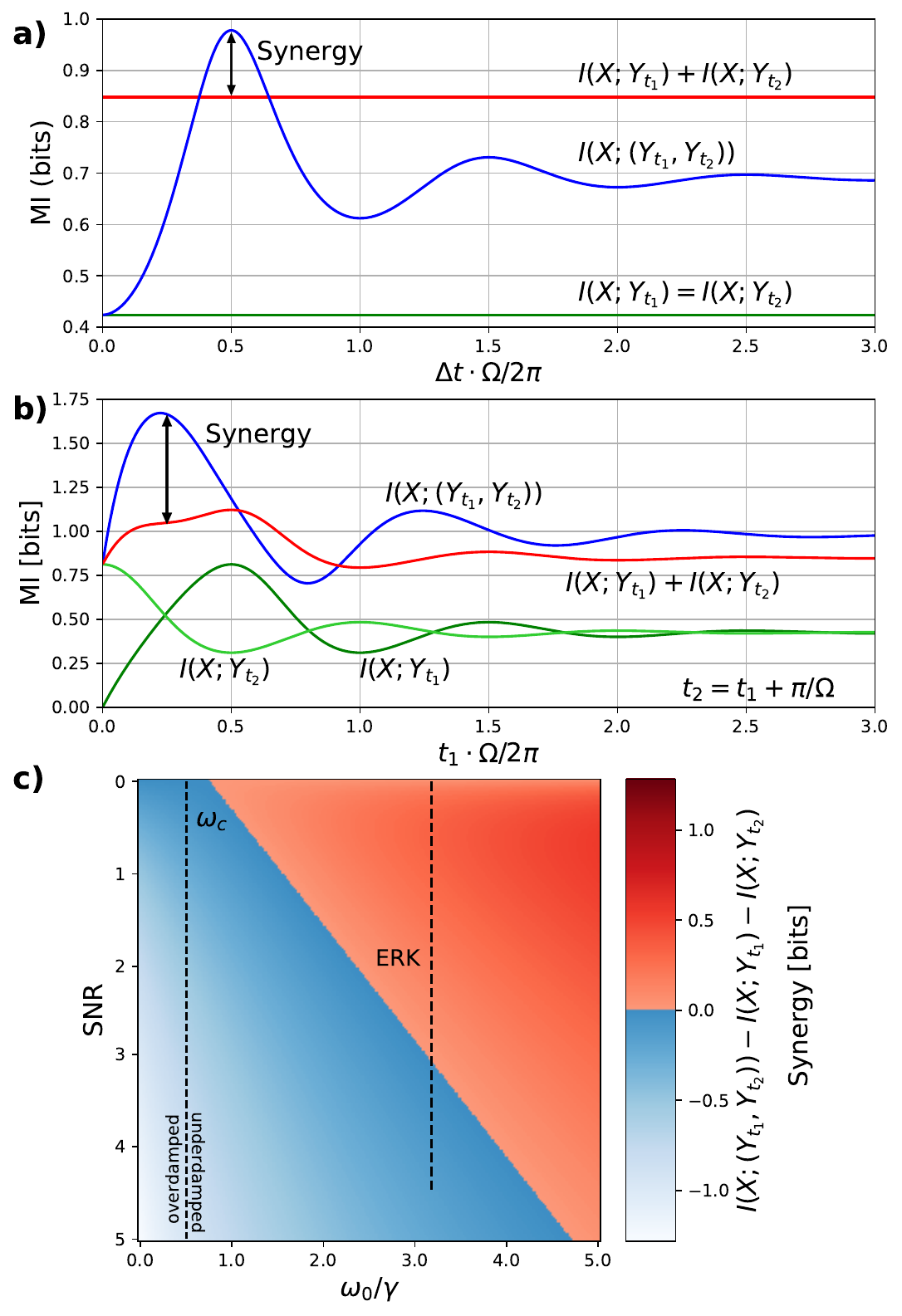}
  \caption{\textbf{Synergy in information transmission between successive measurements}. (a) Information contained jointly in two successive measurements as a function of the delay $\Delta t$ between them, for the oscillatory dynamics of \eqref{eq_un_EOM} at steady state. Synergy is observed when $I(X;(Y_{t_1},Y_{t_2}))>I(X;Y_{t_1})+I(X;Y_{t_2})$.
    (b) Information of two measurements as a function of the time of the first measurement, $t_1$, for fixed delay $\Delta t=t_2-t_1{\ch= \pi/\Omega}$.
    (c) Phase diagram of the synergy in steady state as a function of ${\rm SNR}=\omega_0^2\sigma_x^2/D\gamma$ and $\omega_0/\gamma$, {\ch with $\Delta t=t_2-t_1= \pi/\omega$}. Dashed line shows the range of experimental values estimated from \cite{Dessauges2022} where ERK is activated by optogenetics.
All data is for $\sigma_x^2=1$, $\gamma=1$, and in (a) and (b), $D=5$, $\omega_0=2$. }
\label{fig2}
\end{figure}

Using the same approach as above (see Appendix~\ref{AppendixStaticX}), we can calculate $I(X;Y_{t_1})$ as \eqref{eq:SNR} with:
\beq
S(X;Y_{t_1})=\frac{\sigma_x^2\[ 1-e^{-\gamma t_1/2}\(\cos \Omega t_1 + \frac{\gamma}{2\Omega}\sin \Omega t_1 \) \]^2}{\mathrm{Var}(Y_{t_1}|X)}, \label{eq_un_MIstatic}
\eeq
with $\mathrm{Var}(Y_t|x)=\frac{D\gamma}{4\Omega^2\omega_0^2}[ 4\Omega^2(1-e^{-\gamma t})+\gamma^2e^{-\gamma t}(\cos 2\Omega t-1)-2\gamma\Omega\sin 2\Omega t ]$.
Unlike in the overdamped case, the MI does not necessarily increase with $t_1$, but instead is itself subject to oscillations, and is maximal for $t_1=\pi/\Omega$ (Fig.~\ref{fig2}b).

We can next calculate the mutual information between $x$ and the trajectory of $y$ by introducing the auxiliary variable $z=\dot y$ to make the system Markovian. $I(X;{\bf Y})=I(X;{\bf Y,Z})$ is given by \eqref{eq:SNR} with
    \begin{align}
    S(X;{\bf Y})=&\frac{\sigma_x^2\[ 1-e^{-\frac{\gamma t_1}{2}}\(\cos \Omega t_1 + \frac{\gamma}{2\Omega}\sin \Omega t_1 \) \]^2}{\mathrm{Var}(Y_{t_1})|X)}\nonumber\\
                 &+\frac{\sigma_x^2\omega_0^4e^{-\gamma t_1}\sin^2 \Omega t_1 }{\mathrm{Var}(Z_{t_1}|X)\Omega^2}+\frac{\omega_0^4\sigma_x^2}{2{\gamma^2}D}T, \label{eq_MI_Gauss_under}
    \end{align}
with $\mathrm{Var}(Z_t|x)=\frac{D\gamma}{4\Omega^2}[ 4\Omega^2(1-e^{-\gamma t})+\gamma^2e^{-\gamma t}(\cos 2\Omega t-1)+2\gamma\Omega\sin 2\Omega t ]$.
Note that the $T\to 0$ limit corresponds to the information given by an instantaneous measurement of $Y$ and its derivative $Z$, which gives more information than $Y$ alone  \eqref{eq_un_MIstatic}.

The scaling with observation time $T$  in \eqref{eq_MI_Gauss_under} has the same property of diminishing return as the overdamped case. However, synergy can emerge if we consider two measurements $y_{t_1}$ and $y_{t_2}$. The corresponding mutual information $I(X;Y_{t_1},Y_{t_2})$ may be calculated analytically using the Markovian propagator $p(y_{t_2}|y_{t_1},z_{t_1},x)$ (see Appendix~\ref{AppendixStaticX}). At steady state ($t_1\to\infty$), its simplifies to \eqref{eq:SNR} with:
\begin{align}
S&=\frac{\omega_0^2\sigma_x^2}{\gamma D} +\frac{\sigma_x^2\( 1-e^{-\frac{\gamma}{2} \Delta t}(\cos\Omega \Delta t +\frac{\gamma}{2\Omega}\sin\Omega \Delta t) \)^2}{\mathrm{Var}(Y(\Delta t)|X)+\frac{D}{\gamma\Omega^2}e^{-\gamma \Delta t}\sin^2\Omega \Delta t},
\end{align}
with $\Delta t=t_2-t_1$, which is plotted in Fig.~\ref{fig2}a. We observe that the joint information is maximal when the second measurement is done $\Delta t=\pi/\Omega$ after the first, i.e. at opposite phase. The resting position of the oscillator is then approximately the average of the two measurements, irrespective of the phase of the first measurement. In that case, the two measurements {\ch are} {\em synergistic},
{\ch
${\rm Synergy}=I(X;(Y_{t_1},Y_{t_2}))-I(X;Y_{t_1})-I(X;Y_{t_2})>0$.
They} provide more information together than the sum of each, which are confounded by lack of phase information.
Fig.~\ref{fig2}b shows the more realistic case of a first measurement in finite time, and optimal delay $\Delta t= \pi/\Omega$, confirming that synergy is a generic outcome.
{\ch The measurement times that maximize information transmission are then $t_1^*<\pi/\Omega$ and $t_2^*-t_1^*=\pi/\Omega$, versus $t^*=\pi/\Omega$ for a single measurement.}
The phase diagram of {\ch the steady-state} synergy as a function of the dimensionless parameters of the dynamics, $\omega_0/\gamma$, and the steady-state ${\rm SNR}\equiv \omega_0^2\sigma_x^2/D\gamma$ (Fig.~\ref{fig2}c) shows that synergy emerges when damping is weak and noise is large, which is the regime in which resonant effects are strong.

We can apply our formulas to experimental measurements of the response of the ERK pathway activated by an optogenetic actuator \cite{Dessauges2022} in the oscillatory regime. From Fig.~2e of \cite{Dessauges2022} we estimate $\Omega^{-1}\approx 1.43$ min and $\gamma^{-1}=4.5$ min, so that $\omega_0/\gamma\approx 3.2$, and $\mathrm{Var}(Y_{t\to\infty}|X)\approx 0.013$ in the experiment's arbitrary units of normalized fluorescence. The output variance depends on the dynamic range of inputs, but is lower than $0.06$ in those units, giving a SNR varying between 0 to $\approx 4.5$. In this experimental regime, the total information of two time points may be as large as 2.6 bits, with synergy appearing for $\mathrm{SNR}\lesssim 3.3$ (Fig.~\ref{fig2}c, dashed line). This suggests that synergy may be relevant in the physiological regime of those experiments, and could be exploited by cells in downstream signaling.

So far we have considered the case of an input affecting the resting position of the output $y$. However, other encodings are also common in biological systems, such as frequency modulation, proposed for Msn2 \cite{Hao2012}, Crz1 \cite{Cai2008} and Ca$^{2+}$ \cite{Berridge1998,Boulware2008}, or damping modulation, similar to the distinct dynamical responses of NF-$\kappa$B when stimulated with TNF$\alpha$ or LPS \cite{Covert2005}.
These encodings are no longer Gaussian, so we must turn to numerical methods to estimate mutual information. We first generate solutions to \eqref{eq_un_EOM} for a large number ($N\sim 10^4$) of sampled inputs. We then either calculate the empirical mean of $-\log [P(x|{\bf y})/P(x)]$ using the Markovian expression \eqref{eq_Bayes}, or
use the k-nearest-neighbor (knn) estimators developped by Kraskov, St{\"o}gbauer and Grassberger (KSG) \cite{Kraskov2004}, and Selimkhanov \emph{et al.} \cite{Selimkhanov2014}. While these estimators are very flexible and widely used, they have been criticized for not taking information encoded in the temporal order of successive measurements into account \cite{Tang2021}. We benchmark the estimators using the exact results derived in the previous section and (see Appendix~\ref{AppendixNumerics} for more details).

\begin{figure}[t!]
\centering
  \includegraphics[width=1\linewidth]{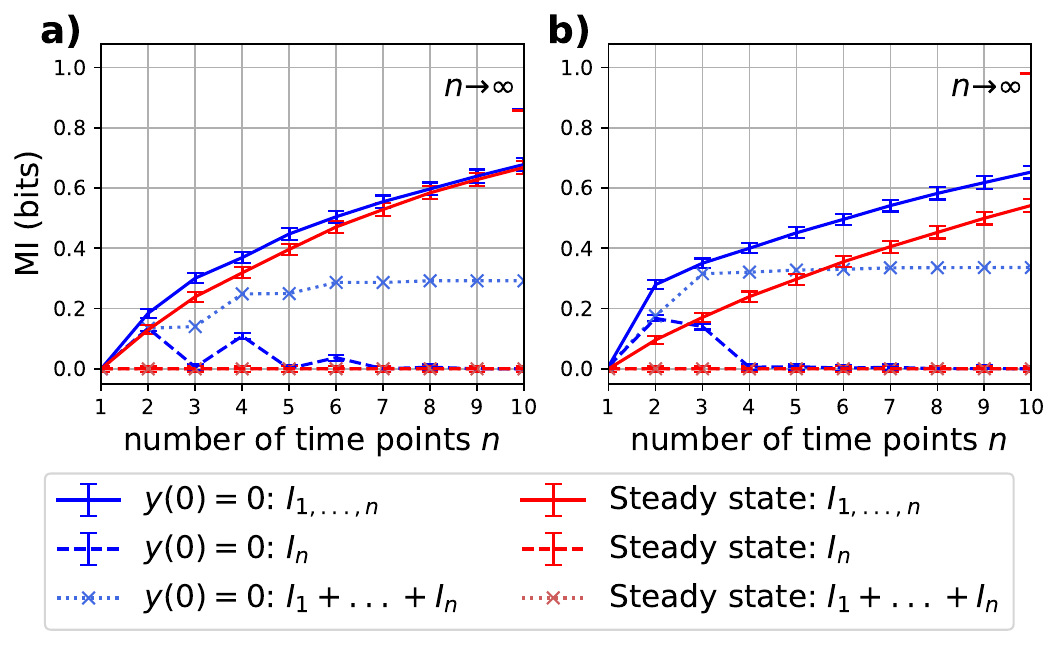}
  \caption{\textbf{Frequency and damping modulation}. Information contained in $n$ successive equidistant measurements, with fixed ($y(0)=0$) or random (steady-state) initial conditions. The binary input consists of (a) two different frequencies in the underdamped regime; and (b) two different damping coefficients in the overdamped and underdamped regimes respectively. 
Solid curves show the joint MI of $n$ measurements, dashed curves the MI of the $n$th measurement alone, and the dotted curves the sum of the individual MIs up to that point. Synergy is observed when the solid line is above the dotted line. Note that since the input is binary, all MI $\leq 1$ bit. The curves were obtained using $T=8$, $D=4$, $\sigma_x^2=1$, and (a) $\omega_0=2$, $\gamma_1=1$, $\gamma_2=6$, {\ch $\Delta t = \pi/\Omega$}, (b) $\gamma=1$, $\omega_{0,1}=2$, $\omega_{0,2}=3$, {\ch $\Delta t = 2\pi/(\Omega_1+\Omega_2)$}.}
\label{fig3}
\end{figure}

We first study frequency modulation by considering an underdamped system \eqref{eq_un_EOM} with null resting position, and two equiprobable input frequencies $\{\omega_{0,1},\omega_{0,2}\}$, {\ch corresponding to two discrete stimuli that the cell is trying to distinguish, and which control the response frequency}.
We initialize the system either at a random value drawn from the steady state, or from $y(0)=0$, and compute the MI in the relaxation dynamics, i.e. we ask how well the two frequencies can be distinguished. When starting from a random value, a single measurement should contain very little information, but two measurements should together allow for a good estimate of $\omega_0$, so we expect to find synergy. These predictions are confirmed by Fig.~\ref{fig3}a, {\ch which shows that information carried by several measurements is larger than the sum of individual ones (global synergy, $I_{1:n}>I_1+\ldots+I_n$)}. In the case of a fixed initial condition, one measurement can already distinguish frequencies since the initial phase is fixed, unless the noise has had time to randomize the phase.

Next, we consider damping modulation by computing the mutual information between an output evolving according to \eqref{eq_un_EOM} with null resting position and  equiprobable binary input damping coefficients $\{\gamma_{1},\gamma_{2}\}$, chosen on both sides of the critical damping transition $\gamma_1>\gamma_c=2\omega_0>\gamma_2$. Once again, synergy is observed, especially when the initial condition is drawn from the steady state (Fig.~\ref{fig3}b). Since all other parameters are equal, single measurements give almost no information on whether the dynamics are over- or underdamped.

In this paper we have derived analytical and numerical solutions for the information carried by an output signal in response to a constant input, which corresponds to the typical experiments of Refs.~\cite{Selimkhanov2014,Potter2017,Granados2018,Tang2021}. By contrast, previous theoretical work on information from temporal trajectories have focused on the case of Gaussian fluctuating inputs at steady state \cite{Tostevin2009,Tostevin2010}, where informations can be decomposed in the frequency domain \cite{Fano1961,Pinsker1964}. For completeness, here we present exact results for a fluctuating input $x$ (following an Ornstein-Uhlenbeck process $\dot x_t=-x_t/\tau_x+\sqrt{2D_x}\eta_x(t)$ with $\<\eta_x(t)\eta_x(t')=\delta(t-t')$) but with a finite observation window $(0,T)$ at steady state.
When $y$ responds to $x$ as before, \eqref{eq_y_over_EOM} or \eqref{eq_un_EOM}, the joint distribution of ${\bf x}=(x_t)_{t\in(0,T)}$ and $({\bf y,z})=(y_t,z_t)_{t\in(0,T)}$ are multivariate Gaussians whose covariance matrices have a tridiagonal structure. After calculating their determinants we obtain exact but lengthy expressions for $I({\bf X};{\bf Y})$, which are given in Appendix~\ref{AppendixDynamicX}. Taking the large time limit gives back the classical information rate of \cite{Tostevin2010}:
\beq
\lim_{T\to\infty}\frac{I({\bf X};{\bf Y})}{T}= \frac{1}{2\tau_x}\left( \sqrt{1+\frac{D_x\tau_x^2}{D\tau^2}}-1 \right),
\eeq
for both the overdamped and underdamped cases.

{\ch Information synergy had been previously discussed in neuroscience as a property of groups of neurons \cite{Schneidman2003}, or between spikes of the same neuron \cite{Brenner2000a}. Our models demonstrate that synergy could also be relevant in cellular signalling in the physiological regime when the response function is dynamic and nonlinear, allowing cells to extract more information from stimuli and to make faster decisions.} Potential candidates for synergetic signalling include Msn2, Crz1 or Ca$^{2+}$, which have been found to use frequency modulation to encode information, and NF-$\kappa$B in response to TNF$\alpha$ and LPS, which can show both over- and underdamped dynamics depending on the stimulus \cite{Covert2005}. Frequency modulation has also been hypothesized to be used in some pathways to encode information \cite{Berridge1998,Boulware2008,Hao2012}.
{\ch Whether such synergistic signaling designs are evolutionary adaptive and how the nature and statistics of the input modulate synergy remain open questions.}
Another outcome of our work is exact solutions for the information content in finite trajectories, which we used to benchmark the performance of mutual information estimators that are frequently used in experimental research. Our results provide analytical foundations for a better understanding of dynamical information in biochemical signalling, and suggest new directions for both experimental and theoretical research.

\emph{Acknowledgments.}
We thank M. Kramar and Huy Tran for fruitful discussions.

\appendix
\onecolumngrid

\renewcommand{\thefigure}{S\arabic{figure}}
\setcounter{figure}{0}

\section{Gaussian mutual informations}\label{Gaussian}

The mutual information between the random vectors $\mathbf X$ and $\mathbf Y$ is given by
\beq
I(\mathbf X; \mathbf Y)= H(\mathbf X)-H(\mathbf X|\mathbf Y)=H(\mathbf X)+H(\mathbf Y)-H(\mathbf X,\mathbf Y), \label{eq_MI_entropies}
\eeq
where $H$ is the (Shannon or differential) entropy. For a multivariate Gaussian distribution with covariance matrix $\mathbf\Sigma$, $H_G=1/2 \log[ \det (2\pi e  \mathbf\Sigma)]$ and thus we obtain for jointly Gaussian $\mathbf X$ and $\mathbf Y$
\beq
I(\mathbf X;\mathbf Y)=\frac{1}{2}\log\( \frac{\det \mathbf \Sigma_{xx}}{\det \mathbf \Sigma_{x|y}} \)=\frac{1}{2}\log\( \frac{\det\mathbf \Sigma_{xx}\det \mathbf \Sigma_{yy}}{\det \mathbf \Sigma} \), \label{eq_mi_GaussianXY}
\eeq
with the covariance matrices $\mathbf \Sigma_{xx}$ of $\mathbf X$, $\mathbf \Sigma_{x|y}$ of $\mathbf X$ given $\mathbf Y$, and 
\beq
\mathbf \Sigma = \begin{bmatrix} \mathbf \Sigma_{xx} & \mathbf \Sigma_{xy}\\
\mathbf \Sigma_{yx}&\mathbf \Sigma_{yy}
\end{bmatrix}
\eeq
the covariance matrix of the joint distribution of $(\mathbf X,\mathbf Y)$ \cite{Tostevin2010,Tkacik2011}.

\section{Mutual information in trajectories: static input}\label{AppendixStaticX}
Here, we present the details of our calculations of the mutual information (MI) in trajectories from the main text when the input is static and drawn from a Gaussian distribution
\begin{align}
p(x)=\frac{1}{\sqrt{2 \pi \sigma_x^2}} e^{-\frac{(x-x_0)^2}{2 \sigma_x^2}}
\label{stimulus}
\end{align}
with mean $x_0$ and variance $\sigma_x^2$. For the output, we first consider overdamped dynamics and then repeat the calculations for underdamped dynamics.

\emph{Overdamped dynamics.--}
The output dynamics are given by
\beq
\dot{y}_t=\tau^{-1}\left(x- y_t \right)+ \sqrt{2D}\eta(t),\label{eq_y_ov_EOM_appendix}
\eeq
where $\av{\eta(t)}=0$ and $\av{\eta(t) \eta(t')}=\delta(t-t')$. The input determines the resting position $x$ of the oscillator. The general solution of this stochastic differential equation with initial condition $y(0)=y_0$ is
\beq
y(t)= y_0e^{-t/\tau} +x (1-e^{-t/\tau}) +\sqrt{2D} \int_0^{t} \eta(t') e^{-(t-t')/\tau} d t'
\eeq
from which we calculate the mean
\beq
\langle y \rangle (y_0,x,t)= y_0e^{-t/\tau} +x (1-e^{-t/\tau}) \label{eq_ov_mean}
\eeq
and the variance of $Y$ given $X=x$
\begin{align}
\mathrm{Var}(Y(t)|X)&= D\tau \left ( 1-e^{-2t/\tau} \right). \label{eq_ov_var} 
\end{align}
For $t\to 0$, $\mathrm{Var}(Y(t)|X)\to 2Dt$ and thus the infinitesimal propagator is
\beq
p(y_{t+\delta t}|y_t,x)= \frac{1}{\sqrt{4 \pi D \delta t}}e^{-\frac{\left[y_{t+\delta t} -y_t-\delta t/\tau[x-y_t]\right]^2}{4D\delta t}}. \label{eq_ov_infprop}
\eeq
Thus $\{Y_t\}$ is a Gaussian process and at any time $t'$ the output will have a Gaussian distribution. For given $x$, the steady state is then
\beq
p_{\text{ss}}(y|x)=\frac{1}{\sqrt{2 \pi D\tau}} e^{-\frac{(y-x)^2}{2 D\tau}} \label{app_eq_ov_ss}
\eeq

We now calculate the mutual information (i) between $X$ and an instantaneous measurement of $Y$, (ii) between $X$ and a time trace $\{Y_t\}$, (iii) between $X$ and $N$ measurements of $Y$ and (iv) finally discuss non-Gaussian inputs.

(i) We begin by calculating the mutual information between $X$ and an instantaneous measurement of $Y$ at time $t_1>0$ when $y(t=0)=y_0=0$. We calculate
\begin{align}
p(x|y_{t_1})&=\frac{p(x)p(y_{t_1}|x)}{p(y_{t_1})}=\frac{1}{Z}e^{-\frac{(x-x_0)^2}{2\sigma_x^2}-\frac{(y_{t_1}-\langle y \rangle (y_0,x,t))^2}{2 \mathrm{Var}(Y(t_1)|X)}} \\
&\overset{!}{=} \frac{1}{Z'}\exp\(-\frac{1}{2}\alpha x^2+\beta x+\gamma \)
\end{align}
where $Z$ and $Z'$ are normalization factors independent of $x$. By inserting eqs.~\eqref{eq_ov_mean} and \eqref{eq_ov_var}, and comparing coefficients, we obtain $\mathrm{Var}(X|Y(t_1))$, the variance of $p(x|y_{t_1})$,
\beq
\frac{1}{\mathrm{Var}(X|Y(t_1))}=\alpha=\frac{1}{\sigma_x^2}+\frac{1}{D\tau}\frac{\(1-e^{-t_1/\tau}\)^2}{1-e^{-2t_1/\tau}}.
\eeq
We can then compute the mutual information from the entropies of $p(x)$ and $p(x|y_{t_1})$, using $H_G=1/2 \log[ 2\pi e  \sigma^2]$ for the entropy of a Gaussian distribution,
\beq
I(X;Y)=\frac{1}{2}\log\(\frac{\sigma_x^2}{\mathrm{Var}(X|Y(t_1))}\)=\frac{1}{2}\log\(1+\frac{\sigma_x^2}{D\tau}\frac{\(1-e^{-t_1/\tau}\)^2}{1-e^{-2t_1/\tau}}\)
\eeq

(ii) Next, we calculate the information between $X$ and a trajectory $\{Y_t\}$ of length $T$ which is measured starting from time $t_1$, again with initial conditions $y(0)=0$. We start by sampling the output trajectory at $N=T/\delta t$ times $\{t_1,t_1+\delta t,t_1+2\delta t,...,t_1+T\}$ to obtain a Gaussian random vector $\mathbf Y=\{Y_{t_1},Y_{t_1+\delta t},...,Y_{t_1+T}\}$. Using the infinitesimal propagator in eq.~\eqref{eq_ov_infprop}, we can calculate
\beq
p(\mathbf y|x)=p(y_{t_1}|x)\prod_t p(y_{t+\delta t}|y_t,x),
\eeq
which, in the limit $\delta t\to 0$, will give the probability of the trajectory $\{y_t\}$. 

Since $p(x)$ and $p(\mathbf y|x)$ are Gaussian, so is
\begin{align}
p(x|\mathbf y)&=\frac{p(x)p(\mathbf y|x)}{p(\mathbf y)} = \frac{p(x)p(y_{t_1}|x)\Pi_t p(y_{t+\delta t}|y_t,x)}{p(\mathbf y)} \\
&=\frac{1}{Z} e^{-\sum_{t}\frac{\left[y_{t+\delta t} -y_t-\delta t/\tau[x-y_t]\right]^2}{4D\delta t}}e^{-\frac{(x-x_0)^2}{2 \sigma_x^2}} e^{-\frac{(y_{t_1}-\langle y \rangle (y_0=0,x,t))^2}{2 \mathrm{Var}(Y(t_1)|X)}}.
\end{align}
As before, we determine the variance of $p(x|\mathbf y)$ by comparing coefficients in the exponent:
\beq
\frac{1}{\mathrm{Var}(X|\mathbf Y)}=\frac{1}{\sigma^2_x}+\frac{1}{D\tau}+\sum_t \frac{\delta t}{2 D \tau^2}=\frac{1}{\sigma^2_x}+\frac{1}{D\tau}+\frac{T}{2 D \tau^2}.
\eeq
This then yields the result
\begin{align}
I(X;\{Y_t\}) =\lim_{\delta t\to 0}I(X;\mathbf Y) = \frac{1}{2} \log \left[1+\frac{ \sigma_x^2}{ D \tau}\left(\frac{T}{2\tau}+(1-e^{-t_1/\tau})^2\right)\right]. \label{app_MI_over_t1}
\end{align}

Repeating the same calculation in steady state, i.e. using eq.~\eqref{app_eq_ov_ss} instead of $p(y_{t_1}|x)$, we obtain
\beq
I(X;\{Y_t\})=\frac{1}{2} \log \left[1+\frac{\sigma_x^2}{ D \tau}\left(1+\frac{T}{2\tau}\right)\right], \label{eq_ov_MItraj}
\eeq
which indeed is the limit of eq.~\eqref{app_MI_over_t1} in the limit $t_1\to\infty$.

(iii) Let us now consider $N$ measurements of $Y$ with $\Delta t$ between two measurements. The probability of measuring $\mathbf y=(y_{t_1},y_{t_1+\Delta t},...,y_{t_1+(N-1)\Delta t})$ is
\beq
p(\mathbf y|x)= p(y_{t_1}|x)\prod_{n=1}^{N-1}\frac{1}{Z}\exp\(-\frac{(y_{t_n}-\langle y\rangle(y_{t_{n-1}},x,\Delta t))^2}{2\mathrm{Var}(Y(\Delta t)|X)}\),
\eeq
where we use the finite-time propagator obtained from eqs.~\eqref{eq_ov_mean} and \eqref{eq_ov_var}. We proceed as in (ii) to obtain the mutual information
\beq
I(X;\mathbf Y)=\frac{1}{2} \log \left[1+\frac{\sigma_x^2}{ D \tau}\( 1+(N-1)\frac{(1-e^{-\Delta t/\tau})^2}{1-e^{-2\Delta t/\tau}} \)\right].
\eeq
In the limit $N\to\infty$, $\Delta t\to 0$, $N\Delta t=T=\text{const.}$, we recover eq.~\eqref{eq_ov_MItraj}.

(iv) We now discuss the case of non-Gaussian input distributions in steady state. In this case, $p(\mathbf y|x)$ is still Gaussian but in principle not the joint distribution with the input $x$. However for the uniform distribution $p(x)=1/L$ for $x\in [0,L]$ it remains Gaussian with 
\beq
\frac{1}{\mathrm{Var}(X| Y)}=\frac{T}{2\tau^2D}+\frac{1}{D\tau}.
\eeq
The entropy of the input is now $H=\log(L)$. The above calculation still holds and we obtain
\begin{align}
I(X;\{Y_t\})&=\log L-\frac{1}{2}\log \left[{2 \pi e D \tau \frac{1}{1+\frac{T}{2\tau}}} \right] \\
&=\frac{1}{2}\log{\left[{\frac{L^2}{2\pi e D \tau}\left(1+\frac{T}{2\tau} \right)}\right]}.
\end{align}

In general, for a non-constant input distribution $p(x)$ we need to make an approximation that the input distribution is well peaked around the mean $\bar{x}$ and expand the non-Gaussian distribution around the peak of the distribution:
\beq
p(x)\approx\frac{1}{Z}\exp\( -\sum_n \partial^n_x (\ln p(x)) (x-\bar{x})^n/n! \)
\eeq
and then truncate at $n=2$. Thus, we can approximate $p(x|\mathbf y)$ as a Gaussian with variance
\beq
\frac{1}{\mathrm{Var}(X| Y)}=\partial_x^2 \ln p(x)|_{\bar x}+\frac{1}{D\tau}+\frac{T}{2 D \tau^2}.
\eeq
Assuming that $p(x|\mathbf y)\approx\delta(x-\bar x)$ is strongly peaked, we can approximate $H(X)$ by
\beq
H(X)=-\int d\mathbf y p(\mathbf y) \int dx p(x|\mathbf y) \log p(x) \approx -\log p(\bar{x})
\eeq
which altogether gives us the mutual information
\begin{align}
I(\{Y_t\};X)&=-\log p(\bar x) +\frac{1}{2}\log{\left[ \frac{1}{2\pi e}\left( \frac{1}{D\tau} +\frac{T}{2 D \tau^2} + \partial_x^2 \ln p(x)|_{\bar x} \right) \right]} =\frac{1}{2}\log\left[ \frac{1}{2\pi e p(\bar x)^2}\left( \frac{1}{D\tau} +\frac{T}{2 D \tau^2} + \partial_x^2 \ln p(x)|_{\bar x} \right) \right],
\end{align}

\emph{Underdamped dynamics.--}
The calculations above are essentially the same when considering underdamped dynamics 
\beq
\ddot{y}_t = -\gamma \dot y_t-\omega_0^2(y_t-x)+\sqrt{2D}\gamma\eta_y(t), \label{eq_un_EOMy}
\eeq
or, as a system of first order equations,
\begin{align}
\frac{d}{dt}\begin{bmatrix} y\\z \end{bmatrix}=\begin{bmatrix} 0 & 1\\ -\omega_0^2 & -\gamma \end{bmatrix}\begin{bmatrix} y\\z \end{bmatrix}+  \begin{bmatrix}0\\ \omega_0^2 x+\sqrt{2D}\gamma\eta_y \end{bmatrix}, \label{eq_un_EOMyz}
\end{align}
where $z_t=\dot y_t$. First, we solve eq.~\eqref{eq_un_EOMyz} for initial conditions $y(0)=y_0$, $z(0)=z_0$:
\begin{align}
y(t)&= x+e^{-\frac{\gamma}{2}t}\( (y_0-x)\cos\Omega t+\[\frac{z_0}{\Omega}+\frac{\gamma}{2\Omega}(y_0-x)\]\sin\Omega t \) + \frac{\sqrt{2D}\gamma}{\Omega}\int_0^td\tau\,\eta(\tau)e^{-\frac{\gamma}{2}\tau}\sin\Omega\tau, \\
z(t)&= e^{-\frac{\gamma}{2}t}\( z_0\cos\Omega t-\[\frac{\gamma z_0}{2\Omega}+\frac{\omega_0^2}{\Omega}(y_0-x)\]\sin\Omega t \) + \frac{\sqrt{2D}\gamma}{\Omega^2}\int_0^td\tau\,\eta(\tau)e^{-\frac{\gamma}{2}\tau}\(\cos\Omega\tau-\frac{\gamma}{2\Omega}\sin\Omega\tau \),
\end{align}
with $\Omega=\sqrt{\omega_0^2-\gamma^2/4}$,from which we can calculate the variances
\begin{align}
&\mathrm{Var}(Y(t))=\sigma_x^2\(1-e^{-\frac{\gamma}{2}t}\(\cos\Omega t+\frac{\gamma}{2\Omega}\sin\Omega t \)\)^2 + \mathrm{Var}(Y(t)|X) \\
& \mathrm{Var}(Y(t)|X)=\frac{D\gamma}{4\Omega^2\omega_0^2}\big( 4\Omega^2\(1-e^{-\gamma t}\) +\gamma^2e^{-\gamma t}(\cos 2\Omega t-1)-2\gamma\Omega\sin 2\Omega t \big), \label{eq_varyx_underdamped_app}\\
&\mathrm{Var}(Z(t))=\sigma_x^2 \frac{\omega_0^4}{\Omega^2}e^{-\gamma t}\sin^2\Omega t + \mathrm{Var}(Z(t)|X) \\
&\mathrm{Var}(Z(t)|X)=\frac{D\gamma}{4\Omega^2}\big( 4\Omega^2\(1-e^{-\gamma t}\) +\gamma^2e^{-\gamma t}(\cos 2\Omega t-1)+2\gamma\Omega\sin 2\Omega t \big). \label{eq_varzx_underdamped_app}
\end{align}
We note the limits
\begin{align}
\mathrm{Var}(Y(t))&\xrightarrow[]{t\to 0} 2D\gamma^2t^3/3 \\
\mathrm{Var}(Y(t))&\xrightarrow[]{t\to \infty} \sigma_x^2+D\gamma/\omega_0^2 \\
\mathrm{Var}(Z(t))&\xrightarrow[]{t\to 0} 2D\gamma^2t \\
\mathrm{Var}(Z(t))&\xrightarrow[]{t\to \infty} D\gamma.
\end{align}
and also compute 
\beq
\mathrm{Cov}(X,Y(t))=\sigma_x^2\( 1-e^{-\frac{\gamma}{2}t}\(\cos\Omega t+\frac{\gamma}{2\Omega}\sin\Omega t\) \) \xrightarrow[]{t\to \infty} \sigma_x^2, \\
\eeq
such that the steady state covariance matrix of $p(y,x)$ is
\beq
\mathbf \Sigma = \begin{bmatrix} \Sigma_{xx} & \Sigma_{xy}\\
\Sigma_{yx}&\Sigma_{yy}
\end{bmatrix} 
= \begin{bmatrix} \sigma_x^2 & \sigma_x^2 \\
\sigma_x^2 & \sigma_x^2+D\gamma/\omega_0^2
\end{bmatrix}.
\eeq

(i) With fixed initial conditions, $y(t=0)=y_0=0$, $z(t=0)=z_0=0$, we calculate
\beq
p(x|y_{t_1})=\frac{p(y_{t_1}|x)p(x)}{p(y_{t_1})}=\frac{1}{Z} e^{-\frac{(x-x_0)^2}{2\sigma_x^2}}e^{-\frac{1}{2\mathrm{Var}(Y(t_1)|X)}\(y_{t_1}-\langle y \rangle (t_1)\)^2}
\eeq
where $\mathrm{Var}(Y(t_1)|X)$ is given by eq.~\eqref{eq_varyx_underdamped_app} and 
\beq
\langle y\rangle (t_1)= x\(1- e^{-\frac{\gamma}{2}t_1}\( \cos\Omega t_1+\frac{\gamma}{2\Omega}\sin\Omega t_1 \) \). \label{eq_meany_underdamped_app}
\eeq
Again, we read off the variance and compute
\beq
I(X;Y)=\frac{1}{2}\log\( 1+\frac{4\omega_0^2\Omega^2\sigma_x^2}{D\gamma}\frac{\(1- e^{-\frac{\gamma}{2}t_1}\( \cos\Omega t_1+\frac{\gamma}{2\Omega}\sin\Omega t_1 \) \)^2}{4\Omega^2(1-e^{-\gamma t_1}) +\gamma^2e^{-\gamma t_1}(\cos2\Omega t_1-1)-2\gamma\Omega e^{-\gamma t_1}\sin 2\Omega t_1 } \)
\eeq

(ii) To calculate the information between a static input $X$ and a trace $\{Y_t\}$ as before, we need to calculate $p(x|\mathbf y)$. However, care must be taken since $\{Y_t\}$ is not a Markovian process. Thus, calculating the probability of a trace $p(\mathbf y|x)$ is not feasible since we cannot compute the infinitesimal generator. But $\{Y_t,Z_t\}$ as defined by the dynamics in eq.~\eqref{eq_un_EOMyz} is a Markov process, allowing us to calculate $p(\mathbf y,\mathbf z|x)$ using the generator
\begin{align}
&p(y_{t+\delta t},z_{t+\delta t}|y_{t},z_{t},x)\sim e^{ -\frac{3}{4D\delta t^3}(y_{t+\delta t}-y_t-z_t\delta t)^2} \\
&\qquad\times e^{-\frac{1}{4D\delta t}(z_{t+\delta t}-z_t-\delta t(\omega_0^2(x-y_t)-\gamma z_t))^2 },
\end{align}
given here up to a normalization factor. Thus,
\beq
p(\mathbf y,\mathbf z|x)=p(y_{t_0},z_{t_0})\prod_t p(y_{t+\delta t},z_{t+\delta t}|y_{t},z_{t},x)
\eeq
and
\beq
p(x|\mathbf y,\mathbf z)=\frac{p(y_{t_0},z_{t_0}|x)p(x)\prod_t p(y_{t+\delta t},z_{t+\delta t}|y_{t},z_{t},x)}{p(\mathbf y,\mathbf z)}
\eeq
which is Gaussian with variance $\mathrm{Var}(X|Y(t),Z(t))$ since both $p(x)$ and $p(\mathbf y,\mathbf z|x)$ are Gaussian. Comparing coefficients leads to
\beq
\frac{1}{\mathrm{Var}(X|Y(t_1),Z(t_1))}=\frac{1}{\sigma_{x}^2}+\sum_t \frac{1}{4D\delta t}(\omega_0^2\delta t)^2+\frac{1}{\mathrm{Var}(Y(t_1)|X)}\(1- e^{-\frac{\gamma}{2}t_1}\( \cos\Omega t_1+\frac{\gamma}{2\Omega}\sin\Omega t_1 \) \)+\frac{\omega_0^4 e^{-\gamma t_1}\sin^2 \Omega t_1}{\Omega^2\mathrm{Var}(Z(t_1)|X)}.
\eeq
To get to the information between $X$ and $\{Y_t\}$, we note that $I(X;\{Y_t\})=I(X;\{Y_t,Z_t\})$ since $Z_t$ is just a function of $Y_t$. Therefore,
\begin{align}
I(X;\{Y_t\})&=I(X;\{Y_t,Z_t\})=\lim_{\delta t\to 0}I(X;\mathbf Y,\mathbf Z) \nonumber\\ &=\frac{1}{2} \log \[ 1+ \frac{\omega_0^4\sigma_x^2}{2D}T+\frac{\sigma_x^2}{\mathrm{Var}(Y(t_1)|X)}\(1- e^{-\frac{\gamma}{2}t_1}\( \cos\Omega t_1+\frac{\gamma}{2\Omega}\sin\Omega t_1 \) \)+\frac{\sigma_x^2 \omega_0^4 e^{-\gamma t_1}\sin^2 \Omega t_1}{\Omega^2\mathrm{Var}(Z(t_1)|X)} \].
\end{align}

(iii) Since $\{Y_t\}$ is not Markovian, calculating the MI in $N$ time points would require calculating $p(\mathbf y,\mathbf z|x)$ and then marginalizing over all $z_t$ which is impractical. However, the case of two time points is the most relevant for questions of synergy and only requires marginalizing one variable. Assuming fixed initial conditions and writing $\Delta t=t_2-t_1$,
\begin{align}
&p(y_{t_2},y_{t_1}|x)=\int_{-\infty}^\infty dz_{t_1} p(y_{t_2}|y_{t_1},z_{t_1},x)p(y_{t_1},z_{t_1}|x). \\
&\quad=\frac{1}{Z}\int_{-\infty}^\infty dz_{t_1}e^{- \frac{e^{-\gamma\Delta t}\sin^2(\Omega \Delta t)}{2\Omega^2\mathrm{Var}(Y(\Delta t)|X)}(z_{t_1}-\tilde z)^2 - \frac{1}{2\mathrm{Var}(Y(t_1)|X)}(y_{t_1}-\langle y\rangle (t_1))^2 - \frac{1}{2\mathrm{Var}(Z(t_1)|X)}(z_{t_1}-\langle z\rangle (t_1))^2}
\end{align}
with $\mathrm{Var}(Y(t_1)|X)$, $\mathrm{Var}(Z(t_1)|X)$ and $\langle y\rangle(t_1)$ given by eqs.~\eqref{eq_varyx_underdamped_app}, \eqref{eq_varzx_underdamped_app}, and \eqref{eq_meany_underdamped_app}, respectively, $\langle z\rangle(t_1)=-\frac{\omega_0^2}{\Omega}xe^{-\gamma t_1/2}\sin\Omega t_1$, and
\begin{align}
\tilde z = \Omega\,\frac{ y_{t_2}-x+ e^{-\frac{\gamma}{2}\Delta t}(x-y_{t_1})\(\cos\Omega\Delta t+\frac{\gamma}{2\Omega}\sin\Omega\Delta t\)}{e^{-\frac{\gamma}{2}\Delta t}\sin\Omega\Delta t} \\
\end{align}
such that evaluating the integral and reading off the variance leads to
\begin{align}
I(X;(Y_{t_1},Y_{t_2}))= \frac{1}{2}\log\Bigg[ 1&+\frac{\sigma_x^2}{\mathrm{Var}(Y(t_1)|X)}\(1- e^{-\frac{\gamma}{2}t_1}\( \cos\Omega t_1+\frac{\gamma}{2\Omega}\sin\Omega t_1 \) \)^2  \\ &+\frac{\sigma_x^2\( 1-e^{-\frac{\gamma}{2} \Delta t}(\cos\Omega \Delta t +\frac{\gamma}{2\Omega}\sin\Omega \Delta t) +\frac{\omega_0^2}{\Omega}e^{-\gamma t_1/2}\sin\Omega t_1 \)^2}{\mathrm{Var}(Y(\Delta t)|X)+\frac{e^{-\gamma \Delta t}\sin^2\Omega\Delta t}{\Omega^2}\mathrm{Var}(Z(t_1)|X)} \Bigg].
\end{align}

\section{Mutual information in trajectories: dynamical input}\label{AppendixDynamicX}

In this appendix, we calculate the mutual information between two time traces. We consider the same output dynamics as before -- first over-, then underdamped -- with a decaying input signal. This scenario is particularly relevant since the degradation of signalling molecules is an important element and a ubiquitous phenomenon in cellular signalling.

\emph{Overdamped dynamics.--}
Input and output are governed by the equations:
\begin{align}
\dot{x}_t &= -\tau_x^{-1} x_t + \sqrt{2D_x}\eta_x(t), \label{eq_ov_EOM_x_app} \\
\dot{y}_t &=  \tau^{-1}\left(x_t- y_t \right)+ \sqrt{2D_y}\eta_y(t) \label{eq_ov_EOM_y_app}
\end{align}
with $\av{\eta_i(t)}=0$ and $\av{\eta_i(t) \eta_j(t')}= \delta(t-t')\delta_{ij}$.

We begin by calculating the steady state covariance matrix $\mathbf\Sigma$ by writing eqs.~\eqref{eq_ov_EOM_x_app} and \eqref{eq_ov_EOM_y_app} in matrix form
\begin{align}
\frac{d}{dt}\begin{bmatrix} x\\y \end{bmatrix}=\begin{bmatrix} -1/\tau_x & 0\\ 1/\tau & -1/\tau \end{bmatrix}\begin{bmatrix} x\\y \end{bmatrix}+ \begin{bmatrix} \sqrt{2D_x}\eta_x\\  \sqrt{2D_y}\eta_y \end{bmatrix},
\end{align}
which gives in Fourier space:
\begin{align}
\begin{bmatrix} \hat{x}\\ \hat{y} \end{bmatrix}=\frac{1}{(-i \omega +1/\tau)(-i \omega +1/\tau_x)}
\begin{bmatrix} -i \omega +1/\tau & 0\\ 1/\tau & -i \omega +1/\tau_x \end{bmatrix} \begin{bmatrix}\sqrt{2D_x}\hat\eta_x\\ \sqrt{2D_y}\hat\eta_y \end{bmatrix}.
\end{align}
Then,
\begin{align}
\mathbf\Sigma = \begin{bmatrix} \Sigma_{xx} & \Sigma_{xy}\\
\Sigma_{yx}&\Sigma_{yy}
\end{bmatrix}
= \int \frac{d \omega}{2 \pi}\begin{bmatrix} 
\av{\hat x \hat x^*} & \av{\hat x \hat y^*} \\
\av{\hat y \hat x^*} & \av{\hat y \hat y^*} 
\end{bmatrix}= \int \frac{d \omega}{2 \pi}\begin{bmatrix} 
S_{xx} & S_{xy} \\
S_{yx} & S_{yy}
\end{bmatrix},
\end{align}
for which we obtain
\begin{align}
\Sigma_{xx} &= D_x\tau_x,\\
\Sigma_{xy} &=D_x \frac{\tau_x}{\tau} \frac{\tau}{1+\tau/ \tau_x}, \\
\Sigma_{yy} &=D_x\tau_x\left[ \frac{1}{\left(1+\tau/ \tau_x \right)}+\frac{D\tau}{D_x\tau_x}\right].
\end{align}
From this we can already compute the MI from an instantaneous measurement of $X$ and $Y$ using eq.~\eqref{eq_mi_GaussianXY}, which gives
\begin{align}
I(X;Y) =\frac{1}{2}\log\( \frac{ \frac{1}{(1+\tau/\tau_x)^2} +\frac{D_y\tau}{D_x\tau_x}}{ \frac{1}{1+\tau/\tau_x} +\frac{D_y\tau}{D_x\tau_x}-\( \frac{1}{1+\tau/\tau_x}\)^2}\) = \frac{1}{2}\log\left[ 1+\frac{A \tau\tau_x}{(1+\tau/ \tau_x)^2+A \tau^2} \right] \label{eq_ov_dynX_staticMI}
\end{align}
where $A=D_x/D_y\tau^2$.

By the same approach as in Appendix~\ref{AppendixStaticX}, we can sample the traces $\{X_t\}$ and $\{Y_t\}$ at $N=T/\delta t$ equally spaced points in time and use the infinitesimal generators to calculate the probability of the resulting $N$-dimensional vectors $\mathbf X$ and $\mathbf Y$:
\begin{align}
p(\mathbf x) &=p(x_{t_0})\Pi_t p(x_{t+\delta t}|x_{t}) \\
p(\mathbf y|\mathbf x) &= p(y_{t_0}|x_{t_0})\Pi_t p(y_{t+\delta t}|y_t,x_t)
\end{align}
where
\begin{align}
p(x_{t+\delta t}|x_{t}) &=\frac{1}{Z} e^{-\sum_t \frac{\left(x_{t+\delta t} -x_t+ x_t \delta t/\tau_x \right)^2}{4 D_x\delta t}} \\
p(y_{t+\delta t}|y_t,x_t) &= \frac{1}{Z'} e^{-\sum_t\frac{\left( y_{t+\delta t}-y_t-{\delta t}/{\tau}(x_t-y_t)\right)^2}{4D_y \delta t}}
\end{align}
with normalization factors $Z$, $Z'$. Next, we calculate, up to normalization factors,
\begin{align}
p(\mathbf x|\mathbf y) &= \frac{p(\mathbf y | \mathbf x)p(\mathbf x)}{p(\mathbf y)} = \frac{p(x_{t_0},y_{t_0})\Pi_t [p(x_{t+\delta t}|x_{t})p(y_{t+\delta t}|y_t,x_t)]}{p(\mathbf y)}\\
&\sim e^{-\sum_t\frac{\left( y_{t+\delta t}-y_t-{\delta t}/{\tau}(x_t-y_t)\right)^2}{4D \delta t}} e^{-\frac{\Sigma_{xx}}{2\det\mathbf\Sigma}{\left(y_{t_0}-\frac{\Sigma_{xy}}{\Sigma_{xx}}x_{t_0} \right)^2}} \times e^{-\frac{x^2_{t_0}}{2\Sigma_{xx}}}e^{-\sum_t\frac{ \left(x_{t+\delta t} -x_t+ x_t \delta t/\tau_x \right)^2}{4 D_x\delta t}} \\
&\overset{!}{=} e^{-\frac{1}{2}\sum_{t t'}C_{t t'} x_t x_{t'}}.
\end{align}
By comparing coefficients we obtain the tridiagonal matrix $C_{t t'}=a_0\delta_{t,t_0}\delta_{t',t_0}+a_1\delta_{t,t_0+T}\delta_{t',t_0+T}+a'\delta_{t,t_0+T>t'>t_0}+b (\delta_{t', t+1}+ \delta_{t',t-1})$ where we use the notation
\begin{align}
A&=\frac{D_x}{D_y\tau^2},\\
a&= \frac{1}{2D_x \delta t} \left[ 1+ ( \delta t/\tau_x -1)^2\right], \\
a'&=\frac{1}{2D_x \delta t} A \delta t^2+a, \\
b&=  \frac{1}{2D_x \delta t}  ( \delta t/ \tau_x -1), \\
\alpha_0&=\frac{1}{\Sigma_{xx}} \left( \frac{\Sigma^2_{xy}}{\Sigma_{xx} \Sigma_{yy}-\Sigma^2_{xy}} +1\right)=\frac{1}{D_x \tau_x}\left[ 1+\frac{A \tau\tau_x}{(1+\tau/ \tau_x)^2+A \tau^2}\right], \\
a_0&=\frac{1}{2D_x \delta t} \left[ \alpha_0 (2D_x \delta t)+ A \delta t^2+b^2 (2D_x \delta t)^2 \right], \\
a_1&= \frac{1}{2D_x \delta t},
\end{align}
or, more graphically:
\begin{align}
C= \left( \begin{array}{cccccc}
a_0 & b & 0&\hdots &\hdots&0\\
b & a' & b & \ddots& &\vdots\\
0&b& \ddots & \ddots & \ddots &\vdots\\
\vdots&\ddots&\ddots & \ddots& b & 0  \\
\vdots& & \ddots & b&a'& b \\
0& \hdots&\hdots &0& b & a_1 \end{array} \right)\label{eq_C_matrix}.
\end{align}

The structure of the matrix for the input $C^{\text{in}}$ is the same,
\beq
p(\mathbf x) =p(x_{t_0})\Pi_t p(x_{t+\delta t}|x_{t}) \overset{!}{=} e^{-\frac{1}{2}\sum_{t t'}C^{\text{in}}_{t t'} x_t x_{t'}},
\eeq
except that $A=0$, i.e. $C^{\text{in}}=C^(A=0)$. Therefore, the next step is to calculate $\det C$ so that finally
\beq
I(\mathbf X;\mathbf Y)=\frac{1}{2}\log\( \frac{\det C}{\det C(A=0)}\).
\eeq
This lengthy calculation is done in Appendix~\ref{AppendixDeterminant}. After taking the limit $\delta t\to 0$, we obtain
\begin{align}
I(\{X_t\};\{Y_t\})= -\frac{ T}{2\tau_x}+ \frac{1}{2}\log  \Bigg( D_x\alpha_0\tau_x  \cosh{\left( \sqrt{1+A\tau_x^2} T/\tau_x \right)}   + \frac{2D_x\alpha_0\tau_x+A\tau_x^2}{2\sqrt{1+A\tau_x^2}}   \sinh{\left( \sqrt{1+A\tau_x^2} T/\tau_x \right)} \Bigg) . \label{eq_ov_MIdynamic}
\end{align}

In the limit $T\to 0$, we recover the static MI (see eq.~\eqref{eq_ov_dynX_staticMI}):
\beq
I(X;Y)=\frac{1}{2}\log\[D_x\alpha_0\tau_x\]= \frac{1}{2}\log\left[ 1+\frac{A \tau\tau_x}{(1+\tau/ \tau_x)^2+A \tau^2} \right].
\eeq
Futhermore, taking the limit $T\to\infty$ yields
\beq
I(\{X_t\};\{Y_t\}) \approx \frac{ T}{2\tau_x}\left( \sqrt{1+A\tau_x^2}-1 \right) \approx \frac{A\tau_xT}{4}=R_oT
\eeq
from which we can read the asymptotic MI rate
\beq
R_o=\frac{A\tau_x}{4}=\frac{\tau_x}{4\tau^2}\frac{D_x}{D_y}. \label{eq_ov_MIrate}
\eeq
In this limit, we should obtain the same result using the approach of Tostevin and ten Wolde for infinite trajectories \cite{Tostevin2009},
\begin{align}
R=- \frac{1}{4 \pi} \int_{-\infty}^{\infty} d \omega \log \left[ 1- \frac{\abs{S_{xy}}^2}{\abs{S_{xx}}\abs{S_{yy}}}\right] =\frac{1}{4 \pi} \int_{-\infty}^{\infty} d \omega \log \left[ 1+\frac{\Sigma(\omega)}{N(\omega)}\right],
\end{align}
where $N(\omega)$ is the power spectrum of the noise (the conditional distribution $P(\mathbf y|x)$), $\Sigma(\omega)=\abs{S_{xy}}^2/\abs{S_{xx}}$ the transmitted input, and $\abs{S_{yy}}=\Sigma(\omega)+ N(\omega)$. We calculate
\begin{align}
N(\omega)&=\frac{ 2D_y \left( \omega^2 +\tau_x^{-2}\right)}{\left(\omega^2+{\tau_x}^{-2} \right)\left(\omega^2+{\tau}^{-2} \right)},\\
\Sigma(\omega)&=\frac{2 D_x \left({1}/{\tau} \right)^2}{\left[ \left(\omega^2+{\tau_x}^{-2} \right)\left(\omega^2+{\tau}^{-2} \right)\right]}
\end{align}
and evaluate the integral
\begin{align}
R=\frac{1}{4 \pi} \int_{-\infty}^{\infty} d \omega \log \left[1+\frac{  \left({1}/{\tau} \right)^2 D_x/D}{\left( \omega^2 +\tau_x^{-2}\right)}\right] =\frac{1}{4 \pi} \int_{-\infty}^{\infty} d \omega \frac{ \left({1}/{\tau} \right)^2 D_x/D}{\left( \omega^2 +\tau_x^{-2}\right)}=\frac{  \tau_xD_x/D}{4\tau^2 }=\frac{A\tau_x}{4}
\end{align}
and indeed recover our previous result from eq.~\eqref{eq_ov_MIrate}.

\emph{Underdamped dynamics.--}
When considering an underdamped system,
\begin{align}
\dot{x}_t &= -\tau_x^{-1} x_t + \sqrt{2D_x}\eta_x(t),  \\
\ddot{y}_t &= -\gamma \dot y_t-\omega_0^2(y_t-x)+\sqrt{2D_y}\gamma\eta_y(t), 
\end{align}
the previous calculation is in principle identical. The steady-state covariance matrix reads
\beq
\mathbf \Sigma = \begin{bmatrix} \Sigma_{xx} & \Sigma_{xy} & \Sigma_{xz}\\
\Sigma_{yx}&\Sigma_{yy} & \Sigma_{yz} \\
\Sigma_{zx}&\Sigma_{zy} & \Sigma_{zz} 
\end{bmatrix} ,
\eeq
where
\begin{align}
\Sigma_{xx}&= D_x\tau_x \\
\Sigma_{xy}&=\Sigma_{yx}= D_x/\tau_x-\frac{2D_x\gamma\omega_0^2}{\(\omega_0^2+\tau_x^{-2}\)^2-\gamma^2\tau_x^{-2}} \\
\Sigma_{xz}&=\Sigma_{zx}= \frac{D_x\omega_0^2}{2(\omega_0^2+\tau_x^{-2}+\gamma\tau_x^{-1}) }\\
\Sigma_{yy}&= \frac{D_y\gamma}{\omega_0^2}+2D_x\(\frac{\omega_0^2}{\tau_x^{-2}+\omega_0^2-\tau_x^{-1}\gamma}\)^2  \( \frac{\gamma^2-2\gamma\tau_x^{-1}+\tau_x^{-2}+\omega_0^2}{2\gamma\omega_0^2}-\frac{2\gamma}{\omega_0^2+\tau_x^{-2}+\tau_x^{-1}\gamma}+\frac{\tau_x}{2} \) \\
\Sigma_{yz}&=\Sigma_{zy}= 0 \\
\Sigma_{zz}&= D_y\gamma+2D_x\(\frac{\omega_0^2}{\tau_x^{-2}+\omega_0^2-\tau_x^{-1}\gamma}\)^2 \( \frac{1}{2\tau_x}-\frac{2(\tau_x^{-2}-\omega_0^2+\tau_x^{-1}\gamma)}{\tau_x^{-1}+\tau_x\omega_0^2+\gamma} + \frac{5\gamma^2\tau_x^{-2}-8\gamma\tau_x^{-1}\omega_0^2+4\omega_0^4+4\tau_x^{-2}\Omega^2}{8\gamma\omega_0^2} \).
\end{align}
This can be used to calculate the MI from an instantaneous measurement using eq.~\eqref{eq_mi_GaussianXY} (which we do not write explicitly for the sake of brevity) and
\beq
\alpha_0=\frac{\Sigma_{yy}\Sigma_{zz}-\Sigma_{yz}^2}{\det \mathbf\Sigma}.
\eeq
By writing the infinitesimal propagator, we obtain
\beq
A = \frac{D_x\omega_0^4}{D_y\gamma^2}.
\eeq
The rest of the calculation is then identical to the overdamped case leading to eq.~\eqref{eq_ov_MIdynamic} with the new values of $A$ and $\alpha_0$. Therefore, in the underdamped case, the asymptotic MI rate is equal to
\beq
R_u=\frac{A\tau_x}{4}= \frac{D_x}{D_y}\frac{\tau_x\omega_0^4}{4\gamma^2}
\eeq

\section{Calculation of the determinant}\label{AppendixDeterminant}
We calculate the determinant of the tridiagonal matrix $C$ in eq.~\eqref{eq_C_matrix}. To begin with,
\beq
\det C=a_0(a_1D_{N-2}-b^2D_{N-3})-b^2(a_1D_{N-3}-b^2D_{N-4}),
\eeq
where $D_L$ is the determinant of a symmetric tridiagonal matrix of size $L\times L$ with on-diagonal terms $a'$ and off-diagonal $b$. We calculate $D_L$ recursively:
\beq
D_L=a'D_{L-1}-b^2 D_{L-2}.
\eeq
Assuming a solution of the form $D_L\sim r^L$, we find
\beq
D_L=\alpha r_1^L+\beta r_2^L,
\eeq
with
$r_{1,2}=\frac{1}{2}\left(a' \pm \sqrt{a'^2-4 b^2}\right)$, with $\alpha+\beta=1$ and $\alpha r_1 +\beta r_2 =a$. Solving for $\alpha,\beta=\frac{1}{2}\left[ 1\pm\kappa \right]$ (with $\kappa=a'/\sqrt{a'^2-4b^2}$):
\beq
D_L=\frac{1}{2} \left[r_1^L+r_2^L+\kappa\left(r_1^L - r_2^L\right) \right], 
\eeq
from which we get
\begin{align}
\det C= \frac{1}{2}\Big[ (a_0 r_1-b^2)(a_1 r_1-b^2)(1+\kappa)r_1^{N-4}  +(a_0 r_2-b^2)(a_1 r_2-b^2)(1-\kappa)r_2^{N-4} \Big]. \label{eq_detC_expanded}
\end{align}

We now successively calculate the terms appearing in eq.~\eqref{eq_detC_expanded}. By keeping only the lowest orders in $\delta t$, we obtain:
\begin{align}
\kappa &\approx \frac{A\delta t^2+2-2\delta t/\tau_x+(\delta t/\tau_x)^2}{\delta t/\tau_x(2-\delta t/\tau_x) \sqrt{1+A\tau_x^2}} \\
r_{1/2}&\approx \frac{1}{2D_x\delta t}\left( 1+\left( \pm\sqrt{1+A\tau_x^2}-1 \right)\delta t/\tau_x \right), \\
a_0r_{1/2}-b^2&\approx \left( \frac{1}{2D_x\delta t} \right)^2\left[ 2D_x\alpha_0\tau_x -1 \pm \sqrt{1+A\tau_x^2}\right]\delta t/\tau_x, \\
a_1r_{1/2}-b^2&\approx \left( \frac{1}{2D_x\delta t} \right)^2\[ \pm\sqrt{1+A\tau_x^2}+1 \]\delta t/\tau_x.
\end{align}
With $N=T/\delta t$ and $\lim_{n\to 0}(1+xn)^{1/n}=e^x$ we obtain
\begin{align}
r_{1/2}^N &= \left( \frac{1}{2D_x\delta t} \right)^N \left( 1+\left( \pm\sqrt{1+A\tau_x^2}-1 \right)\delta t/\tau_x \right)^N \\
&\xrightarrow[\delta t\to 0]{} \left( \frac{1}{2D_x\delta t} \right)^N e^{\left(\pm\sqrt{1+A\tau_x^2}-1 \right) T/\tau_x}
\end{align}

Putting everything together, we evaluate eq.~\eqref{eq_detC_expanded} to obtain
\begin{align}
\det C&\approx \left( \frac{1}{2D_x\delta t} \right)^N  \(\frac{\delta t}{\tau_x^2}\)^2 e^{- T/\tau_x} \Bigg[ \left[ 2D_x\alpha_0\tau_x+A\tau_x^2+\kappa\sqrt{1+A\tau_x^2}2D_x\alpha_0\tau_x \right] \cosh{\left( \sqrt{1+A\tau_x^2} T/\tau_x \right)}  \\ & \qquad+   \left[ \sqrt{1+A\tau_x^2}2D_x\alpha_0\tau_x + \kappa\(  2D_x\alpha_0\tau_x+A\tau_x^2 \) \right] \sinh{\left( \sqrt{1+A\tau_x^2} T/\tau_x \right)} \Bigg],
\end{align}
and by setting $A=0$ we obtain the determinant of the input:
\beq
\det C(A=0) = 2 \left( \frac{1}{2D_x\delta t} \right)^N \(\frac{\delta t}{\tau_x}\)^2 \(1+\kappa(A=0)\).
\eeq

\section{Numerics}\label{AppendixNumerics}
To look for synergy effects beyond the Gaussian approximation that we have used in our analytical calculations, we have numerically simulated the systems discussed in the main text and Appendices~\ref{AppendixStaticX} and \ref{AppendixDynamicX} and used mutual information (MI) estimators from the literature to compute the MI from our samples. However, numerically estimating mutual information (MI) is a challenging problem and universally unbiased MI estimators have been proven not to exist \cite{Holmes2019}. We have therefore implemented and compared several different estimators and use the systems for which we have exact results as benchmarks. In this Appendix, we briefly introduce the various estimators and discuss their performance on our benchmark tests.

\emph{Estimators.--}
Among the mutual information estimators from the literature, $k$-th nearest neighbor (knn) methods are among the most widely used. These estimators use the geometric distribution of the samples in the joint space of all possible inputs and outputs to locally estimate the probability density $p(x)$ of a point $x$ by assuming that the density is uniform around $x$ on a sphere encompassing its $k$ nearest neighbors. We have implemented the two original versions proposed by Kraskov, St{\"o}gbauer and Grassberger \cite{Kraskov2004} (which we abbreviate KSG 1 and 2). Since these algorithms use a geometric representation of the samples, one must define a metric by which to measure the distance between two points. We chose to use both the metrics induced by the $\ell_2$ and the $\ell_\infty$ norms.

Another knn estimator has been proposed by Selimkhanov \emph{et al.} \cite{Selimkhanov2014}. The Selimkhanov estimator only requires binning the input distribution whereas the KSG estimators requires no binning at all. Importantly, this allows the KSG estimator to be applied to systems with dynamic input, since binning becomes impractical for a high-dimensional input. However, to use the KSG estimator for discrete input distributions, it is necessary to add some small noise to the inputs in order to break their degeneracy \cite{Kraskov2004}, which may adversely affect the quality of the results. The Selimkhanov estimator can do without this.

Disadvantages of knn-based estimators include bad performance for samples whose underlying distribution is locally non-uniform and the fact that the number of samples needed for convergence increases exponentially with the true value of the MI one tries to estimate \cite{Gao2015}. Another disadvantage when using knn estimators to estimate the MI between random vectors was pointed out by Tang \emph{et al.} \cite{Tang2021}, namely that since knn methods are based solely on the geometrical distribution of the samples, they are insensitive to the order of the elements of the vector. These estimators will thus ignore information encoded in the (in our case temporal) order of the data points. Tang \emph{et al.} circumvent this by training Markov models to estimate the probability of each trace individually. In our case, however, this is not necessary since we have the generative model at hand and can use the infinitesimal propagators to calculate the probability of each sampled trace directly. We proceed as in \cite{Selimkhanov2014} and \cite{Tang2021}, further details can be found in the supplementary materials of these articles.

First, we calculate the entropy $H(\{Y_t\}_i|x_i)$ of the output given $x_i$, one of $M$ inputs with probability $q_i$. If the input distribution is continuous, this will require binning. Then, using the $N_i$ samples $\mathbf y^{ij}=\{y^{ij}_{t_0},y^{ij}_{t_1},...y^{ij}_{t_0+T}\}$ ($j=1,...,N_i$) generated under input $x_i$, we calculate
\beq
H(\mathbf Y^i|x_i)=-\langle \log p(\mathbf y^{ij}|x_i)\rangle \approx -\frac{1}{N_i}\sum_{j=1}^{N_i}\log p(\mathbf y^{ij}|x_i), \label{eq_entropyestimate}
\eeq
where we compute the probability using
\beq
p(\mathbf y^{ij}|x_i)=p(y_{t_0}^{ij}|x_i)\prod_m p(y^{ij}_{t_{m+1}}|y^{ij}_{t_m},x_i),
\eeq
as explained in Appendix~\ref{AppendixStaticX}, either by using the infinitesimal propagator if we wish to compute the MI in the entire trace, or the finite time propagator of we aim to calculate the MI in a finite number of measurements. From this, we can compute the total conditional entropy
\begin{align}
H(\mathbf Y|X)&=\sum_{i=1}^Mq_i H(\mathbf Y^i|x_i) \\
&=-\sum_{i=1}^M\frac{q_i}{N_i}\sum_{j=1}^{N_i}\log p(\mathbf y^{ij}|x_i).
\end{align}
Similarly, we write
\beq
p(\mathbf y^{ij})=\sum_{k=1}^Mq_k p(\mathbf y^{ij}|x_k)
\eeq
and use this to estimate the unconditional entropy
\begin{align}
H(\mathbf Y)&=-\sum_{i=1}^M\frac{q_i}{N_i}\sum_{j=1}^{N_i}\log p(\mathbf y^{ij}) \\
&=-\sum_{i=1}^M\frac{q_i}{N_i}\sum_{j=1}^{N_i}\log \( \sum_{k=1}^Mq_k p(\mathbf y^{ij}|x_k) \).
\end{align}
Taken together, this allows us to estimate
\beq
I(X;\mathbf Y)=H(\mathbf Y)-H(\mathbf Y|X).
\eeq

This estimation does not require that the joint distribution of input and output is Gaussian. If this is the case, one may also use the empirical covariance of the samples to directly estimate the MI according to eq.~\eqref{eq_mi_GaussianXY}.

We calculate errors for the KSG estimators using the approach proposed in \cite{Holmes2019}. For the Selimkhanov estimator as well as our own approach, we calculate the empirical variance of each entropy estimate (eq.~\eqref{eq_entropyestimate}) and use standard error propagation to obtain the error of the MI.

\begin{figure*}
\centering
  \includegraphics[width=0.98\linewidth]{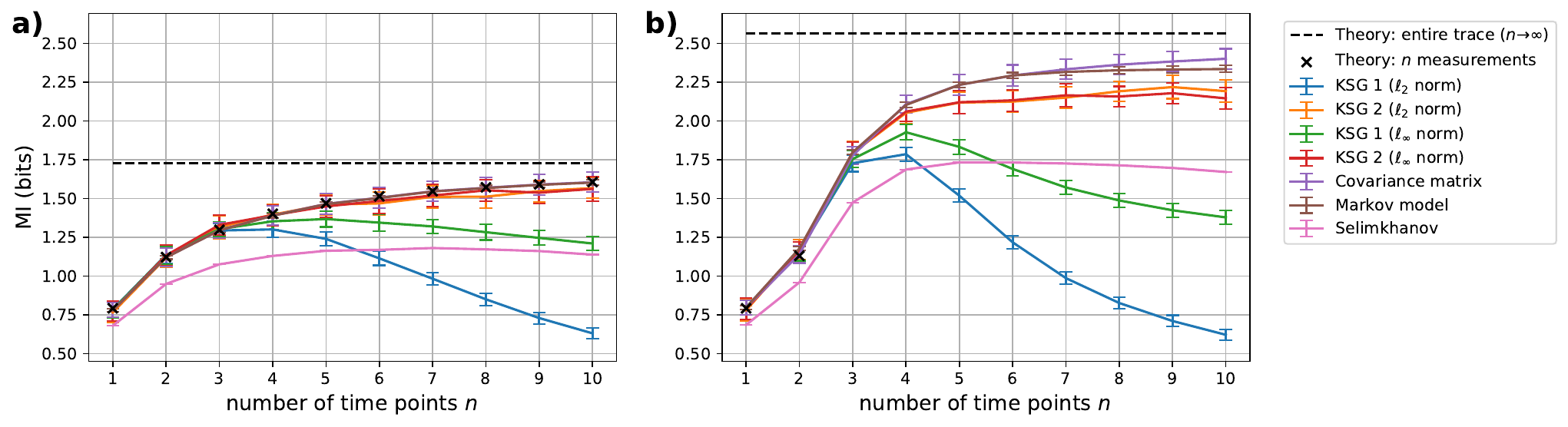}
\caption{\textbf{Benchmarks - static input}. We benchmark the mutual information estimators using the exact results obtained for (a) overdamped and (b) underdamped output dynamics by plotting the MI contained in $n$ equidistant measurements of the output trace, here using $N=20000$ samples and $k=20$. Exact values are shown in black, with the dashed line being the value for $n\to\infty$ (see eqs.~\eqref{eq_MI_Gauss_over} and \eqref{eq_MI_Gauss_under}). We found that all estimators except are accurate for low $n$, but that the Selimkhanov estimator requires much larger $N$ to converge. For increasing $n$, the estimates of KSG 1 and Selimkhanov drop off and do not converge unless greatly increasing the sample size $N$. KSG 2 is more stable independently of whether the $\ell_2$ or $\ell_\infty$ metric is used. However, convergence problems appear in some cases when further increasing $n$ beyond $n \approx 15$. For overdamped dynamics, where we know the exact MI for any $n$, we observe that the covariance matrix-based estimate and our own Markov model-based estimator are the most accurate, although the exact results are within error bars of the KSG 2 estimates. We are unsure why the estimates of KSG 2 differ more strongly from the Markov model and covariance estimators in the case of underdamped dynamics as compared to the overdamped case.} 
\label{fig1app}
\vspace{0.7cm}
  \includegraphics[width=0.98\linewidth]{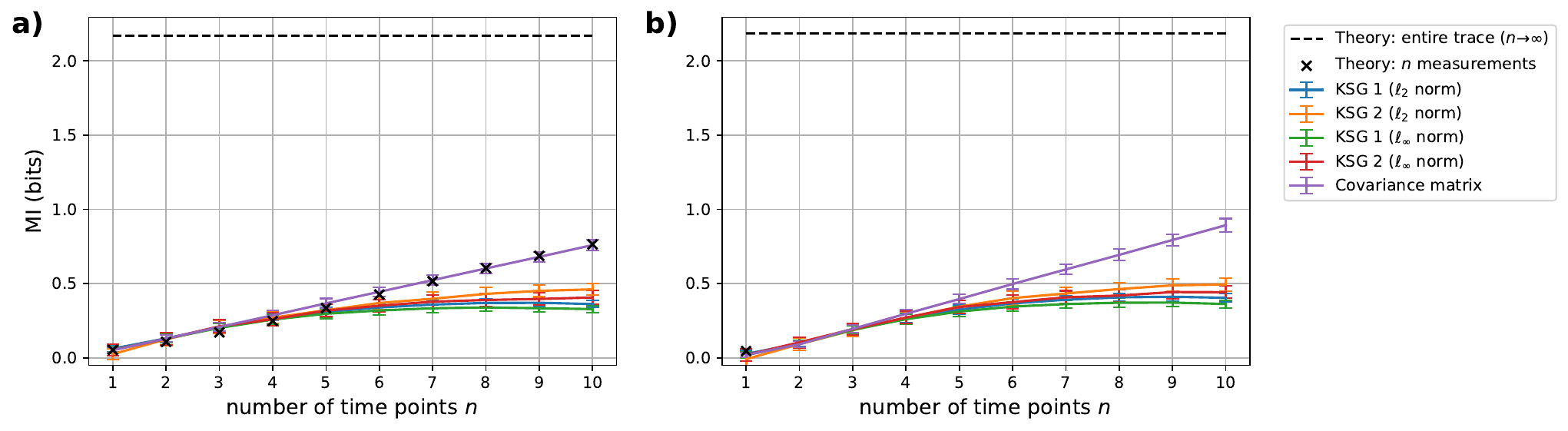}
\caption{\textbf{Benchmarks - dynamic input}. We benchmark the mutual information estimators using the exact results obtained for (a) overdamped and (b) underdamped output dynamics as in Fig.~\ref{fig1app}, using $N=100000$ samples and $k=20$. The dynamic input scenario reveals the limitations of the knn-based estimators. When increasing the dimension of the joint vector space of inputs and outputs, for constant $N$ the sampled points in this space become more sparsely distributed and the performance of the estimator decreases. Neither the Selimkhanov estimator nor our own Markov model-based estimator were used since they require a discrete input distribution and binning the high-dimensional input distribution is impractical.} 
\label{fig2app}
\end{figure*}

\emph{Benchmarks.--}
The exact results derived in Appendices~\ref{AppendixStaticX} and \ref{AppendixDynamicX} allow us to benchmark the performance of the various estimators. We consider first the results from Appendix~\ref{AppendixStaticX}, i.e. a static input drawn from a Gaussian distribution (see Fig.~\ref{fig1app}). Convergence problems appear for the knn estimators when the dimensionality $n$ of the samples increases. This is to be expected since for increasing $n$ the dimension of the joint vector space of inputs and outputs will grow and thus the $N$ available data points will be more sparsely distributed, leading to inaccurate estimates. To compensate this, $N$ would need to be increased further. 

Our own Markov model estimator remains accurate even for high $n$. Indeed, when estimating the MI in the entire trace using the infinitesimal propagator, i.e. $n\gg 1$, we compute in the overdamped case $I(X;\{Y_t\})=1.72\pm0.05$ bits, close to the exact result $I(X;\{Y_t\})=1.73$ bits. In the underdamped case, the estimate gives $I(X;\{Y_t\})=2.55\pm0.05$ bits, versus the exact value $I(X;\{Y_t\})=2.56$ bits. 

When considering dynamic inputs as in Appendix~\ref{AppendixDynamicX}, the limits of the knn estimators when dealing with high-dimensional data become apparent. As can be seen in Fig.~\ref{fig2app}, the KSG algorithms fail to converge for $n>5$ and their predictions grossly underestimate the MI. Our numerical results indicate that for these algorithms to converge it would be necessary to increase the sample size by several orders of magnitude as compared to the static input case, which was not feasible given the available computational resources.

In general, the knn-based estimators prove to be reliable when using low-dimensional data but convergence problems appear in high-dimensional spaces. Our own Markov model-based estimator is more reliable but only applicable when the input distribution can be binned or is already discrete. Directly calculating the MI from the covariance matrix of the joint input-output distribution works well but is restricted to jointly Gaussian processes.

\end{document}